\documentstyle[prd,preprint,aps,epsfig]{revtex}
\tightenlines      
\newcommand{\be}{\begin{equation}}
\newcommand{\ee}{\end{equation}}
\newcommand{\bea}{\begin{eqnarray}}
\newcommand{\eea}{\end{eqnarray}}
\newcommand{\bean}{\begin{eqnarray*}}
\newcommand{\eean}{\end{eqnarray*}}

\newcommand{\gapproxeq}{\lower
.7ex\hbox{$\;\stackrel{\textstyle >}{\sim}\;$}}
\newcommand{\lapproxeq}{\lower
.7ex\hbox{$\;\stackrel{\textstyle <}{\sim}\;$}}

\begin{document}

\title{ A pedagogic model for Deeply Virtual Compton Scattering 
with quark-hadron duality}
\author{Frank E. Close$^1$\thanks{e-mail: F.Close1@physics.ox.ac.uk} 
and Qiang Zhao$^2$\thanks{e-mail: Qiang.Zhao@surrey.ac.uk}}
\address{1) Department of Theoretical Physics,
University of Oxford, \\
Keble Rd., Oxford, OX1 3NP, United Kingdom}%
\address{2) Department of Physics, 
University of Surrey, Guildford, GU2 7XH, United Kingdom}


\maketitle
\begin{abstract}
We show how quark-hadron duality can emerge
for valence spin averaged structure functions, and for
 the non-forward distributions of
Deeply Virtual Compton Scattering. Novel factorisations of the non-forward
amplitudes are proposed. Some implications for large angle scattering 
and deviations from the quark counting rules are illustrated.
\end{abstract}
\vskip 1.cm


\section{Introduction}

There has been much recent interest in two related, though distinct, processes:
(i) Deeply Virtual Compton Scattering
(DVCS), measured in non-forward Compton scattering 
$\gamma^*(q^2) p \to \gamma p$~\cite{Ji,Rad,collins-97,vand-98};
 (ii) Bloom-Gilman duality~\cite{bloom,duality} for the imaginary 
 part of {\it forward} Compton scattering,
where the electroproduction of $N^*$'s at lower energies and
momentum transfers empirically averages smoothly around the scaling curve 
$F_2(W^2,Q^2)$ measured
at large momentum transfers for both proton and neutron targets~\cite{jlab1}. 

In this paper we shall develop a pedagogic model 
of the structure functions for non-diffractive
inelastic scattering from a simple composite system
that satisfies duality. Then we shall investigate its 
implications for the non-forward distribution amplitudes of DVCS~\cite{Ji,Rad}
in certain kinematic regimes, compare with existing models and abstract some 
general features in hope of gaining insight into the physical significance of
 measurements planned at Jefferson Lab and HERA. 

Our point of departure is the work of Refs.~\cite{IJMV,ClIs01},
where the partons (``quark") were treated as spinless constituents.
Reference~\cite{IJMV} analysed 
duality in the context of a large-$N_c$-based
relativistic model with an empirically inspired
 linear potential; it focussed on the dynamics
required for a confined struck parton to behave as though it were free, 
and illustrated
how scaling ensured.  In a complementary approach, Ref.~\cite{ClIs01} 
investigated the circumstances whereby the structure function 
for inelastic scattering
$F(x)$ - whose magnitude in leading twist is in proportion to the
(incoherent) sum of the squares of the
constituent charges - can in general match with
 the excitation of individual resonances which is driven by the
coherently summed square of constituent charges. 
In the present paper we
shall combine these to show how the (non-diffractive, ``valence")
 scaling function emerges for a two body 
state, 
whose spinless constituents have arbitrary charges.
 
This $F(x)$ is intimately related to the imaginary part of the forward
Compton amplitude,  
 data for which lead to the probability
distributions for the partons. Analogously, data on Deeply Virtual Compton
Scattering may be used to
determine non-forward parton distribution amplitudes $F(x,\xi,t)$. 
Here restricting ourselves to a system of two spinless constituents,
we
study the implications of the model for DVCS, and discuss the structure
and physics of $F(x,\xi,t)$. 

The text is arranged as follows. In Sec. II, a simple model 
for a two-body system with arbitarily charged scalar constituents
is introduced; structure functions and sum rules for the forward and non-forward
Compton scattering are derived.  
In Sec. III, we show how the scaling functions
for the forward and non-forward Compton scattering 
arise in this model and note also relations between DVCS and the forward
amplitudes which may be more general than the specific model. 
Functionally the generalised parton distribution
function for the DVCS is discussed for both 
Ji and Radyushkin's frames in Sec. IV. 
Based on the functionally generalised form, a phenomenological
investigation of the non-forward parton distribution in DVCS
is made in Sec. V.
Summary remarks are given in Sec. VI.

\section{A Simple Model}

The original model of Refs.~\cite{IJMV,ClIs01}
ignored spin entirely and considered 
the inelastic scattering of a ``scalar electron" {\it via} exchange of a
``scalar photon" from a composite system of (two) spinless constituents.
The differential cross section was then written as
\be
\label{x-sect}
\frac{d\sigma}{dE^\prime d\Omega_f} = \frac{g^4}{16\pi^2} \frac{E_f}{E_i} 
\frac{1}{Q^4}{\cal F} \ ,
\ee
where the scalar coupling $g$ had dimensions of mass, 
and the factor multiplying
the scalar structure function ${\cal F}$ corresponded to the Mott cross
section.

In this paper we shall consider the more physical situation where
a spin 1/2 
electron scatters by exchanging a vector photon. 
The cross section then has the standard form in terms of two
structure functions $W_1$ and $W_2$, which depend on $\nu$ and $Q^2$
 
\bea
\frac{d\sigma}{dE^\prime d\Omega_f} &=& \sigma_{Mott}[ 
 W_2(\nu,Q^2) + 2W_1(\nu,Q^2)\tan^2\frac{\theta_e}{2}] \nonumber \\
&=& \frac{\sigma_{Mott}}{\epsilon(1+\tau)}[(1-\epsilon)W_1 + 
\epsilon (1+\tau) W_2] \ ,
\eea
where $\tau\equiv\nu^2/Q^2 $, and 
$\epsilon \equiv \rho_{LL} / \rho_{TT}$
is the ratio of longitudinal and transverse density matrix elements.
As before we consider a composite system of two spinless constituents.
While this is still not the real world, the model contains many
of the important physical features: there are analogues of resonances, 
the inelastic structure function exhibits scale invariance 
and ``quark-hadron" duality~\cite{IJMV}. 

Ref.~\cite{IJMV} demonstrated the duality for the case 
of a single particle
in a potential (effectively bound to an infinitely massive 
electrically neutral 
partner). Ref.~\cite{ClIs01} extended some of those ideas by
considering a composite state made of two equal mass
scalars, i.e., ``quarks" $q_1$ and $q_2$ with charges
$e_1$ and $e_2$ respectively at positions $\vec{r}_1$ and $\vec{r}_2$. 
It was shown how
for the imaginary part of the forward Compton amplitude (hence the
inelastic structure function), 
the excitation of resonance states of opposite parity interferes
destructively in all but leading twist. 
In this paper we demonstrate how the model leads 
to scaling behaviour  for the structure functions and then 
 extend it to the imaginary part of
the non-forward Compton amplitude (non-forward distribution
function~\cite{Ji,Rad}). In subsequent section 
some general features for the relation
between the forward and non-forward parton distribution function
will be abstracted and discussed.

\subsection{Forward Compton Scattering \label{fcs}}

 The ground state
wavefunction for the composite state
is $\psi_0(\vec{r})$, where $\vec{r}_{1,2} = \vec{R} \pm \vec{r}/2$
defines the centre of mass and internal spatial degrees of freedom. 
A photon of momentum
$\vec{q}$ is absorbed with an amplitude proportional to
$\sum_i e_i \exp(i \vec{q} \cdot \vec{r}_i)$,
 which excites  a ``resonant" state with
angular momentum $L$, described by the wavefunction $\psi_L(\vec{r})$.

In the previous work which ignored spin entirely~\cite{ClIs01}, 
the inelastic structure function in Eq.~(\ref{x-sect}) 
was given by a sum of squares of transition
form factors weighted by appropriate kinematic factors,

\begin{equation}
{\cal F}(\nu,\vec{q}) = \sum_{N=0}^{max} \frac{1}{4E_0E_N} |F_{0,N}
(\vec{q})|^2 \delta(E_N - E_0 - \nu) \ ,
\label{eq:r3}
\end{equation}
where $\vec{q} \equiv \vec{p_i} - \vec{p_f}$, the form factors
$F_{0,N}$ represent transitions from the ground state to states characterised
by principal quantum number $N(\equiv L + 2k$, where $k$ is 
the radial and $L$ the
orbital quantum numbers). 

Focussing on the internal
coordinate $ \vec{r}$, the transition amplitude is proportional to
\begin{eqnarray}
&(e_1 + e_2) [\exp(i \vec{q} \cdot \vec{r}/2) 
+ \exp(-i \vec{q} \cdot \vec{r}/2)]\nonumber\\
+&(e_1 - e_2) [\exp(i \vec{q} \cdot \vec{r}/2) 
- \exp(-i \vec{q} \cdot \vec{r}/2)]. \nonumber
\end{eqnarray}
The expansion,
$\exp(iqz/2) = \sum_L i^L P_L(\cos \theta) j_L(qr/2) (2L + 1)$,
projects out the even and odd partial waves such that the form factor is
proportional to
\begin{eqnarray}
F_{0,N(L)}(\vec{q}) 
& \sim & \int dr r^2 \psi_L^*(r) \psi_0(r) j_L(qr/2) \nonumber\\
&&\times [(e_1 + e_2) \delta_{L=even} +
(e_1 - e_2) \delta_{L=odd}] \ .
\label{eq:r1}
\end{eqnarray}

The resulting structure function, summed over resonance excitations, will
receive contributions from $L = even \ (odd)$ 
in proportion to $(e_1 \pm e_2)^2$.
Since $N \equiv L+2k$, $L=even \ (odd)$ will imply also that 
$N = even\equiv 2n$ or $N=odd= 2n+1$,
and we can expose these separate contributions to the structure function
(we factor out the charges to make them explicit too),

\begin{equation}
{\cal F}(\nu,\vec{q}) = \sum_{N(n)} 
\frac{1}{4E_0E_N} [F^2_{0,2n}(\vec{q}) (e_1 + e_2)^2 +
F^2_{0,2n+1}(\vec{q}) (e_1 - e_2)^2] \delta(E_N - E_0 - \nu) \ ,
\label{eq:r2}
\end{equation}
or equivalently:
\begin{eqnarray}
{\cal F}(\nu,\vec{q})& =& \sum_{N(n)} \frac{1}{4E_0E_N} 
[(F^2_{0,2n}(\vec{q}) + F^2_{0,2n+1}(\vec{q}))(e_1^2 + e_2^2) +
2e_1e_2(F^2_{0,2n}(\vec{q}) - F^2_{0,2n+1}(\vec{q}))]\nonumber\\
&&\times\delta(E_N - E_0 - \nu) \ .
\label{eq:r2a}
\end{eqnarray}

This simple example exposes the physics rather clearly. The excitation
amplitudes to resonance states contain both diagonal ($e_1^2 + e_2^2)$
 and higher twist
terms ($\pm 2 e_1 e_2$) in the flavour basis. The former set added
constructively for any $L$ and the sum over the complete set of states
can now logically give
  the
deep inelastic curve~\cite{duality}; the latter enter with opposite
phases for even and odd $L$ and destructively interfere. The critical feature
that this exposes is that {\it at least one complete set of resonances of
each symmetry-type has been summed over}. 
In general, the strength of the structure function
 will be  proportional to $e_1^2 + e_2 ^2$
only if the excitation of even and odd $L$ states sum to equal strengths.
Ref.~\cite{ClIs01,IJMV} derived
the circumstances under which this can occur.

Consider scalar quarks confined in a linear potential, 
and described by the Klein-Gordon 
equation such that $V^2(\vec{r}) = \beta^4 r^2$, where $\beta^4$ 
is a generalized,
relativistic string constant (compare also Ref.~\cite{FKR}).
The energy eigenvalues are then $E = \pm E_N$ where
$E_N = \sqrt{2\beta^2(N+3/2) + m^2}$ and $m$ is the mass
of the interacting constituent (``quark"). 
 This choice gives a spectrum that is in accord with that observed empirically
~\cite{FKR,PDG,Cl01rome} since the energy
eigenvalues follow from the similarity to the Schr\"odinger
equation for a non-relativistic harmonic oscillator potential; 
the wave functions
are algebraically as for the non-relativistic case and thereby
enable analytic solutions.

The contribution to ${\cal F}(q)$ from the
$N (\equiv L+2k)$ set of degenerate levels is
\be
F^2_{0,N}(\vec{q}) = \frac{1}{N!} (\frac{\vec{q}^2}{2 \beta^2})^N 
\exp(\frac{-\vec{q}^2}{2 \beta^2}) \ ,
\label{ffactors}
\ee
from which one can immediately
see that $\sum_{N=0}^{\infty} F^2_{0,N}(\vec{q}) = 1$. 
It is interesting to note that any individual contribution,
$F_{0,N}(\vec{q})$, reaches its  maximum value when $\vec{q}^2 = 2 \beta^2 N$,
 at which point $F^2_{0,N} = F^2_{0,N+1}$. 
This coincidence is true for all juxtaposed partial waves at 
their peaks, which gives a rapid approach to the equality
of $\sum_{n=0}^{\infty} F^2_{0,2n}(\vec{q})$ and 
$\sum_{n=0}^{\infty} F^2_{0,2n+1}(\vec{q})$. 

We turn now to the more physical case of real spinning electron and photon. 
For scattering from a system of spinless constituents, 
the leading contributor at large $Q^2$~\cite{su6break}
is the longitudinal response function
\be
\label{response-func}
 R_L \Rightarrow \frac{\nu}{4M^2x^2} [\nu W_2 - 2Mx W_1] \ .
\ee
The resonance sum 
for $ R_L$ is analogous 
to that in Eq.~(\ref{eq:r3}) for ${\cal F}$ but with an extra 
factor of $(E_0 + E_N)^2$.  Hence~\cite{IJV}
\be
R_L(\nu,\vec{q}) = \sum_{N=0}^{max} \frac{1}{4E_0E_N} |f_{0,N}
(\vec{q})|^2 [(E_0 + E_N)^2 \delta(\nu + E_0 - E_N)] \ ,
\ee
where 
\be
|f_{0,N}(\vec{q})|^2 \equiv 
(e_1^2 + e_2^2)[F^2_{0,2n}(\vec{q}) + F^2_{0,2n+1}(\vec{q})] +
2e_1e_2[F^2_{0,2n}(\vec{q}) - F^2_{0,2n+1}(\vec{q})] \ ,
\ee
one has $\nu_{max} < |\vec{q}|$ and the sum over $N$ 
denotes equivalent sum over $n$ for $N=2n$
and $N=2n+1$. 
Furthermore, recall that the energy eigenvalues 
for the Klein-Gordon equation are $E = \pm E_N$ where 
$E_N = \sqrt{2\beta^2(N+3/2) + m^2}$. It will be helpful now 
to take 
$E_N \geq 0$ and to rewrite $R_L$ as
\bea
\label{rl}
R_L(\nu,\vec{q})& =&\sum_{N=0}^{max} \frac{1}{4E_0E_N} |f_{0,N}
(\vec{q})|^2 \nonumber\\
&& \times [(E_0 + E_N)^2 \delta(\nu + E_0 - E_N) - 
(E_0 - E_N)^2 \delta(\nu + E_0 + E_N)] \ .
\eea
This immediately gives the sum rule
\[
S(\vec{q}) \equiv \int_{-\infty}^{+\infty} d\nu R_L(\vec{q},\nu) =
 \sum_{N=0}^{+\infty}|f_{0N}(\vec{q})|^2 \ ,
\]
and hence
\begin{equation}
\int_{-\infty}^{+\infty} d\nu R_L(\nu,\vec{q}) =
 [(e_1^2 + e_2^2) + 2e_1e_2 e^{-(2\vec{q})^2/4 \beta^2}] \ ,
\label{eq:r4}
\end{equation}
which generalizes as 
\begin{equation}
S(\vec{q}) \equiv \int_{-\infty}^{+\infty} d\nu R_L(\nu,\vec{q}) =
 [(e_1^2 + e_2^2) + 2e_1e_2 F_{00}(2\vec{q})] \ .
\label{eq:r10}
\end{equation}
That this is the
correct generalisation rather than $F_{00}^4(\vec{q})$ as 
in Ref.~\cite{ClIs01},
will be obvious from the results of the next Section [Eq.~(\ref{eqsum})].
It also agrees with the physical picture in Fig.~\ref{fig:(1)}, 
where absorption and emission
by different constituents leads to a momentum mismatch in the internal 
wavefunction of $\vec{q} + \vec{k} \to 2\vec{q}$.  
Ref.~\cite{IJMV}
has shown how such a model satisfies scaling and duality 
in the particular case when $e_2 = 0$. Our results enable this to
be generalised to constituents with
arbitrary charges. We shall show this in Sec. III. First we generalise
the above to non-forward Compton scattering amplitudes 
as preparation for our derivation
of the scaling form of DVCS.

\subsection{Non-forward Compton Scattering \label{nfcs}}

We have calculated the imaginary part of the forward Compton 
scattering amplitude 
as a sum over the intermediate coherent ``resonance" states. 
This is experimentally
accessible by measuring $
\frac{d\sigma}{dE'd\Omega_f}(eA \to eX) 
$
 from the target $A$. We now consider the generalisation 
 to the imaginary part of the
non-forward amplitude: $\gamma(q) A \to \gamma(k) A$.
Clearly, when we specialise to the case $\vec{k} \to \vec{q}; t \to 0$,
we must recover our results above.

As before,
focus on the internal
coordinate $ \vec{r}$, and separate the transition amplitude 
into contributions from $N \equiv 2n = even$ and 
$N \equiv 2n+1 = odd$ excited states. 

The generalization is:

\bea
R_L(\nu,\vec{q}, \vec{k},t) &=& \sum_{n(N)} \frac{1}{4E_0E_N} 
(E_0 \pm E_N)^2 \delta (\nu + E_0 \mp E_N) \nonumber \\
&&\times \left(\sum_{L=0(1)}^N
[{\cal F}_{0,2n}^{(L)}(\vec{q}) {\cal F}_{0,2n}^{(L)}(\vec{k})  (e_1 + e_2)^2 +
{\cal F}_{0,2n+1}^{(L)}(\vec{q}) {\cal F}_{0,2n+1}^{(L)}(\vec{k})  (e_1 - e_2)^2]
d^L_{00}(\hat{k} \cdot \hat{q})\right)\ ,
\label{eq:r21}
\eea
where $d^L_{00}(\theta) \equiv \sqrt{\frac{4\pi}{2L+1}} Y_{L0}(\theta)$,
and $\hat{q}$ and $\hat{k}$ denote the unit vector in the direction
of $\vec{q}$ and $\vec{k}$, respectively.
Whereas in the forward case we dealt immediately 
with $\sum_N|F_{0N}(\vec{q})|^2$,
here we must first sum over the various $L$ states within a given $N$. 
Thus (see Appendix 1)
 one now has 
\be
\sum_{N} F_{0N}(\vec{q})F_{N0}(\vec{k}) = \sum_{N} 
\frac{1}{N!}\left(\frac{\vec{q} \cdot \vec{k}}{2\beta^2}\right)^N \exp
(-\frac{\vec{q}^2 + \vec{k}^2}{4\beta^2}) \ ,
\ee
where due to the $\delta$ function in Eq.~(\ref{eq:r21}), 
$t \equiv -|\vec{q} - \vec{k}|^2$. 
Hence the sum over all states gives
\[
(\sum_{N=even }+ \sum_{N= odd})F_{0N}(\vec{q})F_{N0}(\vec{k}) = \exp
(-\frac{(\vec{q}- \vec{k})^2}{4\beta^2}) \equiv F_{00}(|\vec{q} - \vec{k}|) \ ,
\]
while that over opposite phases gives
\[
(\sum_{N=even} - \sum_{N=odd})F_{0N}(\vec{q})F_{N0}(\vec{k}) = \exp
(-\frac{(\vec{q}+ \vec{k})^2}{4\beta^2}) \equiv F_{00}(|\vec{q} + \vec{k}|) \ .
\]

The generalisation to the non-forward case of 
the sum rule Eq.~(\ref{eq:r10}) is 
\be
S(\vec{q},\vec{k}) \equiv \int_{-\infty}^{+\infty} d\nu   R_L(\nu,\vec{q}, \vec{k},t)
= F_{00}(|\vec{q} - \vec{k}|) \left[(e_1^2 + e_2^2)  + 2e_1e_2 
\frac{F_{00}(|\vec{q} + \vec{k}|)}
{F_{00}(|\vec{q} - \vec{k}|)}\right] \ .
\label{eqsum}
\ee
If we restrict attention to $t << Q^2$, in which case
$|\vec{q} - \vec{k}| << |\vec{q} + \vec{k}|$, we have
the sum rule
\[
\int_{-\infty}^{+\infty} d\nu   R_L(\nu,\vec{q}, \vec{k},t) 
= F_{00}(t) (e_1^2+e_2^2) \equiv 
F_{00}(t) \int_{-\infty}^{+\infty} d\nu   R_L(\nu,\vec{q}) \ .
\]
This is the essence of the Ji-Radyushkin sum rule~\cite{Ji,Rad,Ji-Mel-Song}.

For Compton scattering at $90^\circ$, for the case $q^2 \equiv k^2$, we have

\[
\int_{-\infty}^{+\infty} d \nu R_L(\nu,\vec{q}) 
\to (e_1 + e_2)^2 e^{-(\vec{q}^2/2\beta^2)} \to (e_1 + e_2)^2
F_{el}^2(\vec{q}) \ ,
\]
which illustrates how the counting rules of elastic scattering at large 
angles\cite{brodsky} can emerge from duality. An essential part of 
such a derivation is the assumed mass degeneracy between states of
a given $N$ but of different $L$. Generally, 
there will be
$L$-dependent mass shifts within a given $N$ level in the spectrum
(e.g. in a more realistic picture with spin 1/2 $q\overline{q}$, these can be
induced by gluon exchange
and described perturbatively 
 with a Fermi-Breit Hamiltonian). 
These will spoil the exact $L$-degeneracy within any fixed $N$
and lead to
oscillations about a smooth $s^{-n}$ dependence\cite{brodsky} at 90$^\circ$,
where $s$ is the c.m. energy squared. 
Such oscillations are seen in the data\cite{oscil}
It is important to realise that such oscillations at
90$^\circ$ will be a signal primarily for mass splitting $within$ 
a given $N$ level. 
This is distinct from the more usual measure of
duality, as discussed in most of this paper, which involves 
the energy gap $between$ multiplets of different $N$
and different parity
(and would be present even in the absence of gluon exchange mass splittings).

We now establish how scaling obtains in this picture. 

\section{Forward Compton Scattering: Scaling of $F_2(x,Q^2)$ } 

It is not immediately obvious that the above model scales. 
Nor do the detailed
algebraic manipulations that follow easily reveal 
why this property emerges. One can begin to see how the scale invariance arises 
by focussing on the
shape of the curve for an excited state $N$, and in particular the position of
its maximum. This maximum is at $x_{bj} \sim 1/A$ for $A$ constituents,
for any excited state $N$ as we now show.

The scattering is on energy shell whereby $\nu_N = E_N-E_0$. The three
momentum is dispersed, giving a maximum magnitude for the contribution
to the structure function when
 $\vec{q}^2_N = 2 \beta^2 N$. The energy eigenvalues
 $E_N = \sqrt{2\beta^2 (N+3/2) + m^2}$
satisfy $E_N^2 - E_0^2 = 2 \beta^2 N$ and so at the maximum we
have  $E_N^2 - E_0^2 \equiv \vec{q}_N^2$, whereby $\vec{q}_N^2 - 
\nu^2_N \equiv Q_N^2 = (E_N-E_0)2E_0$. Hence the peak occurs when 

\[
Q_N^2 = 2E_0 \nu_N \ .
\]
Thus using the kinematic variable $x_{bj} \equiv Q^2/2M\nu$, 
where $M$ is the mass of the
ground state hadron $\equiv \sum_i E_0^{(i)}$ (i.e., summed over all
constituents), then for excitation of the $N$-th level, the peak occurs at

\[
x_{bj}^{(N)} \equiv \frac{Q^2_N}{2 M \nu_N} = \frac{E_0}{M} \ ,
\]
which suggests that the peak occurs at a common value of $x_{bj}$ for all $N$.
Consequently we find the physically sensible result that the peaking
of the structure function occurs at $E_0/M \sim 1/A$ 
where $A$ is the number of active constituents. 
To obtain the explicit form of the distribution we replace the sum over 
discrete levels by an integral, and use Stirling's formula (for 
details see Sec. V of Ref.~\cite{IJMV}, 
where we generalize their results to the
case of constituents with arbitrary 
charge).

Focus on Eq.~(\ref{rl}) and the leading piece at high $Q^2$, 
i.e., the $(e_1^2 + e_2^2)$ term. Write this, following Ref.~\cite{IJMV} as
\bea 
 R_L(\vec{q},\nu)& =&(e_1^2 + e_2^2) 
\sum_{N=0}^{+\infty} 
 \Delta N \frac{1}{4E_0E_N}
\frac{1}{N!} (\frac{\vec{q}^2}{2 \beta^2})^N 
\exp (-\frac{\vec{q}^2}{2 \beta^2})
\nonumber \\
&&\times [(E_N+E_0)^2 \delta(E_N - E_0 - \nu) \nonumber\\
&&- (E_0 - E_N)^2  \delta(E_N+E_0+\nu)] \ ,
\label{rlforward1}
\eea
where $\Delta N = 1$. 

Some essential steps in manipulating this are 
to rewrite the $\delta$-functions, e.g.
\[
\delta(E_N - (E_0 + \nu)) \to \frac{E_N}{\beta^2} 
\delta\left[(N+\frac{E_0^2}{2\beta^2}) - 
\frac{(\nu+E_0)^2}{2\beta^2}\right] \ ,
\]
then to use Stirling's formula whereby
\[
\frac{1}{N!} \to \frac{1}{\sqrt{2\pi N}} \frac{1}{N^N} e^N \ ,
\]
and to take a continuum limit $\sum_{\Delta N} \to \int d\nu$.
Then tedious algebra gives
\be
 R_L(\vec{q},\nu) = \frac{(\nu + 2E_0)^2}{4 \beta E_0 \sqrt{\pi} \nu} 
 \exp\left[\frac{\nu^2 + 2E_0 \nu}{2 \beta^2}
\ln\left(\frac{Q^2+\nu^2}{\nu^2 + 2E_0 \nu}\right) - \frac{Q^2 + 
\nu^2}{2 \beta^2} + 
\frac{\nu^2 + 2E_0 \nu}{2\beta^2}\right] (e_1^2 + e_2^2)\ ,
\label{nearscaling}
\ee
and a similar term with $\nu \to -\nu$.

As $Q^2 \to \infty$ for fixed
$x_{bj} \equiv \frac{Q^2}{2M\nu}$,
\bea
 R_L(\nu,\vec{q}) \to F_L(x_{bj},Q^2) &=&  
 \frac{Q^2}{8 \beta \sqrt{\pi} M x_{bj}  E_0} e^{-\frac{M^2}{\beta^2}
(\frac{E_0}{M} - x_{bj})^2} (e_1^2+ e_2^2) \nonumber\\
&&\times\left[\theta\left(\frac{Q^2}{2Mx_{bj}}\right) 
- \theta\left(-\frac{Q^2}{2Mx_{bj}}\right)\right] \ .
\label{rlforward2}
\eea
For spinless consituents it is $R_L$ that has been forced to scale.
The transverse response function $R_T$ in this model is suppressed
by $O(1/Q^2)$~\cite{IJV}. This is in agreement with the result for 
spinless partons in the deep inelastic regime.
The implication for the structure function $ W_2$ is that
\be
\label{eq:17}
|\nu |W_2(x_{bj},Q^2) =  F_2(x_{bj})
 =  \frac{x_{bj}^2 M^2}{\beta \sqrt{\pi} E_0} e^{-\frac{M^2}{\beta^2}
(\frac{E_0}{M} - x_{bj})^2} (e_1^2+ e_2^2) \ .
\ee

As discussed in Section~\ref{fcs}, the higher twist term 
vanishes in the limit of $N (n)\to \infty$, where the parity {\it even} and
{\it odd} partial waves are summed to the same strengths, i.e.,
$\sum_{n=0}^{\infty}F^2_{0,2n}(\vec{q})
=\sum_{n=0}^{\infty}F^2_{0,2n+1}(\vec{q})$. 

Equation~({\ref{eq:17}) immediately gives the sum rule:
\be
\int_{-\infty}^{+\infty} \frac{F_2(x_{bj})}{x_{bj}} dx_{bj} 
= e_1^2+e_2^2 \ ,
\ee
which can be compared to Eq.~(\ref{eq:r10}), and is analogous 
to the Gottfried sum rule of the naive parton model~\cite{gottfried}.

The momentum sum rule for the structure function $F_2(x_{bj})$
can be also investigated,
\be
\label{momentum-sum-rule}
\int_{-\infty}^{+\infty} F_2(x_{bj}) dx_{bj}
=\frac{1}{M}(E_0+\frac{\beta^2}{2E_0})(e_1^2+ e_2^2) \ .
\ee
This result sheds light on the ``dynamics" of the duality picture, namely,
the relation between resonance phenomena and naive parton model.
On the one hand, it shows that
the results of the most naive parton model, which is valid at high $Q^2$ and
where the partons can be regarded as quasi-free particles
with distribution $\delta(x-1/A)$, can be only realised at $\beta\to 0$. 
In this case, we find
\be
\int_{-\infty}^{+\infty} F_2(x_{bj}) dx_{bj}
=\frac{E_0}{M} \ ,
\ee
which is the analogue of the familiar sum rule that gives the 
energy fraction carried by the charged constituents.

The structure function $F_2(x_{bj})$ 
exhibits an intuitive connection between resonance excitations
and the naive parton model. 
With $\beta = 0.4$ GeV
for the potential and 
$m = 0.33$ GeV for the constituent mass, we plot $F_2(x_{bj})$ 
versus $x_{bj}$ in Fig.~\ref{fig:(3)} (solid curve); 
this has a broad and flattened behavior and is dominated
by the resonance excitations. 
Obviously,
if $\beta\to 0$, we obtain the Delta function peaking at 
$x_{bj}\to 1/2$ for the equal mass constituents, which recovers
the naive parton model prediction
for the valence contribution to $F_2(x_{bj})$~\cite{su6break}. 
To show this,
we also present the calculation (dashed curve) 
for a weak potential
with $\beta = 0.1$ GeV in contrast to the solid curve, 
where the dashed curve peaking at $x_{bj}=E_0/M$
is clear.
Thus we see how 
 the physics of the naive parton model is recovered as $\beta \to 0$:
in this limit the
constituents are free, all excitation levels are degenerate
and the structure function collapses to an effective expression,
$(e_1^2 + e_2^2)\delta(x_{bj} - 1/A)$.
The physical distribution is smeared around
$x_{bj} = 1/A$; the position of the peak is related 
to the number of active participants
that share momenta, and the width of the peak 
is driven by their momentum spread
which is in turn related to their confinement 
and the energy gap for resonance excitation.

Equation~(\ref{momentum-sum-rule}) suggests that 
the resonance phenomena violate
the naive parton model result. Recalling that the groundstate energy
$E_0=\sqrt{3\beta^2+m^2}$, 
where $3\beta^2 \sim \vec{p}_T^2 + p_z^2$ denotes the Fermi motion
momentum of a constituent, the second term in Eq.~(\ref{momentum-sum-rule})
can be understood as a kinetic energy correction 
to the naive parton model result associated with
the longitudinal component of the Fermi momentum.

Certainly,
this model has many features that prevent it being a serious 
dynamical description of the real world. In particular,
although we have generated a pedagogic picture of how 
the sum over all the resonances leads to a scaling curve
which peaks in a physically sensible region, the quantitative details
of how the scaling function behaves away from this region
are not sensibly produced by the model. For example,
the model allows a kinematic range $-\infty < x <+\infty$, 
whereas the physical region for deep inalastic scattering is limited
to $0\le x\le 1$.

With such caveats in mind, we now investigate the
(unmeasured) non-forward DVCS amplitude in this model, with the
goal of abstracting its general features. 

\section{Nonforward DVCS: Scaling of $F_2(x,\xi,t )$} 

The extensive discussions of non-forward Compton scattering 
in the literature have used two frames. 
One (which we shall refer to as the Radyushkin's kinematics)
has the momenta of the initial state as for forward Compton and throws all 
of the momentum
transfer $\Delta_{\mu}$ 
into the final state. 
Thus
$\gamma(q) + A(P) \to \gamma(k = q - \Delta) + A(P+\Delta)$ where 
$t \equiv - \Delta^2$.
Here we use $A$ to denote the target system which consists
of two spinless constituents.
In this frame, the initial $\gamma A$ system
defines the longitudinal $z$ direction and the final state particles 
have in general transverse
components.
An alternative choice is the symmetric frame of Ji, 
where the transfer $\Delta$ is
shared by initial and final states. Thus
$\gamma(q) + A(P-\Delta/2) \to \gamma(k = q - \Delta) + A(P+\Delta/2)$. 
In this frame, both the initial and final
$\gamma A$ systems have non-zero transverse momenta.

First we define the four vectors as follows
\bea
P &\equiv & [\bar{M};0;0], \nonumber \\
q &\equiv & [\frac{Q^2}{\bar{M} \eta} - \eta \bar{M}; 0; 
-\frac{Q^2}{\bar{M} \eta} - \eta \bar{M}]/2 \ , \nonumber \\
\Delta &\equiv & [0;-\Delta_T; -\xi \bar{M}] \ .
\eea
For Radyushkin's kinematics one has the initial target momentum $p \equiv P$
and hence $\bar{M} \equiv M$. Furthermore $x_{bj} \equiv \eta \sim \xi$. 
For Ji's
kinematics on the other hand, one has $p \equiv (P-\frac{\Delta}{2})$ and 
$\bar{M}^2 \equiv M^2 + \frac{\Delta^2}{4}$. In this frame 
$x_{bj} \equiv \frac{2\xi}{2+\xi}$.

Comparison with our previous calculation is most immediate 
if we use Radyushkin's frame
which has 
the initial $\gamma A$ system
defining the longitudinal $z$ direction. Results are identical 
in Ji's frame, as we
shall illustrate later.

The analogy of Eq.~(\ref{rlforward1}) for nonforward scattering becomes
(see also Appendix 1),
\bea 
R_L(\nu,\vec{q},\vec{k},t)
&=& \sum_{N=0}^{\infty} \Delta N \frac{1}{4E_0E_N}
\frac{1}{N!} (\frac{\vec{q} \cdot \vec{k}}{2 \beta^2})^N 
\exp (-\frac{\vec{q}^2 + \vec{k}^2}{4 \beta^2})(e_1^2 + e_2^2)\nonumber \\
&& \times [(E_N+E_0)^2 
\delta(E_N - E_0 - \nu)-(E_0 - E_N)^2 \delta(E_N+E_0+\nu)] \ .
\label{rldvcs1}
\eea
First replace $k \to q-\Delta$ throughout, so 
we consider $R_L(\nu,q,\Delta,t)$.
Then after tedious algebraic manipulations, analogous to those used 
in Sec. III,
we find a scaling behaviour as follows:
\be
 R_L(\nu,q,\Delta,t) \to F_L(x_{bj},\Delta,t,Q^2) =     
 \frac{Q^2}{8 \beta \sqrt{\pi} Mx_{bj}E_0} 
e^{-\frac{M^2}{\beta^2}
(\frac{E_0}{M} - x_{bj}+\frac{\vec{q} \cdot \vec{\Delta}}{2 M \nu})^2}
e^{-(\frac{\vec{\Delta}^2}{4 \beta^2})}(e_1^2 + e_2^2) \ .
\label{rldvcs2}
\ee
Note that in this model 
$F_{00}(t) \equiv e^{-(\frac{\vec{\Delta}^2}{4 \beta^2})}$ 
and so we see the explicit presence of the
elastic form factor multiplying a skewed distribution function where, 
in effect,
the $x_{bj}$ has been shifted relative to the forward Compton case, 
in a $t$-dependent manner: 
$x_{bj} \to x_{bj}(1- \frac{\vec{q} \cdot \vec{\Delta}}{Q^2})$.

The above was all in Radyushkin's frame. If instead we had 
calculated in Ji's
frame, all the steps follow analogously leading to an identical expression
to the above except that
 the argument of the exponent is modified:
\[
\left(\frac{E_0}{M} - x_{bj} 
+\frac{\vec{q} \cdot \vec{\Delta}}{2 q \cdot p}\right)^2
\rightarrow
\left(\frac{E_0}{M} - x_{bj}+\frac{\vec{q} \cdot \vec{\Delta}}
{2 q \cdot (p - \frac{\Delta}{2})}\right)^2 \ .
\]
Although superficially these
appear to differ,
when expressed in terms of observables they are identical. 

To see most immediately what is happening, we first
ignore corrections of $O(\Delta^2/Q^2)$. 
Then the argument of the exponent becomes
in Radyushkin's kinematics:
\[
\left(\frac{E_0}{M} - x_{bj}+\xi_{Rad}/2\right)^2 \ ,
\]
while for Ji's kinematics it is:
\[
\left[\frac{E_0}{M} - x_{bj}
+\left(\frac{\xi_{Ji}}{2 + \xi_{Ji}}\right)\right]^2 \ .
\]
However, for the imaginary part, which is all that we are considering
here, there are the identities~\cite{rad-prd-98}:
\[
\left(\frac{\xi_{Ji}}{2 + \xi_{Ji}}\right) 
\equiv \xi_{Rad}/2 \equiv x_{bj}/2 \ ,
\]
and so in either frame and in the particular kinematics
$t << Q^2$, we have
\be
F_L \sim e^{-(\frac{E_0}{M} - \frac{x_{bj}}{2})^2} F_{00}(t) \ ,
\label{eq:r16}
\ee
which leads to the amusing factorization:
\be
F_L(x,\xi,t) \to F_L(\frac{x}{2},0,0) F_{el}(t) \ ,
\label{xover2}
\ee
where $x \equiv x_{bj}$. This is in contrast to a commonly used
{\it ansatz} $F(x,\xi,t) \to F(x,0,0) F_{el}(t)$~\cite{vand-98}. 
We shall return to this
later at Eqs.~(\ref{factor}) and (\ref{eq:31}).
This specific example reinforces the result of a recent study~\cite{miller}
which suggested that the factorization form $F(x,0,0) F_{el}(t)$ 
is not a general result
in DVCS.

We now show how this generalises when $\Delta^2/Q^2$ corrections are included.
In this case we need to look more carefully at 
\bea
\frac{q \cdot \Delta}{q \cdot p}= \xi \frac{q_3}{q_0} &=& 
- \xi\left(1 + \eta^2\frac{M^2}{Q^2}\right)
/ \left(1 - \eta^2\frac{M^2}{Q^2}\right)
\nonumber \\
&=& - x_{bj}\left(1 + \eta^2 \frac{M^2}{Q^2}\right)\frac{\xi}{\eta} \ .
\eea
The most general form is to write these in terms of invariants:
\[
\vec{q} \cdot \vec{\Delta} \equiv - q \cdot \Delta 
\equiv (k^2 + Q^2 - \Delta^2)/2 \ .
\]
The symmetry of the ensuing expressions is clearest 
if we define $K^2 \equiv -k^2$. A particular
example is forward Compton scattering for 
which $K^2 \equiv Q^2>0$. Although in practice we will be interested
in $K^2 = 0$, we keep this variable in the formula to give
\[
\frac{\vec{q} \cdot \vec{\Delta}}{2 q \cdot p} 
\equiv \frac{x_{bj}}{2}\left( 1 - \frac{K^2 + \Delta^2}{Q^2}\right) \ , 
\]
which also confirms that $\eta = \xi$ for large $Q^2$. 
We therefore define
\be
x_{bj}^{in} \equiv \frac{Q^2}{2p \cdot q} ; \  
x_{bj}^{fin} \equiv \frac{K^2 + \Delta^2}{2p \cdot q} \ .
\ee
Hence for the forward case, since $t=\Delta^2 \to 0$,
$x_{bj}^{fin} \equiv x_{bj}^{in}$ will be recovered when $K^2 \equiv Q^2$.

Then Eq.~(\ref{rldvcs2}) can be manipulated into the following form
\bea
F_L(x_{bj}^{in},x_{bj}^{fin},t,Q^2) & =&   (e_1^2 + e_2^2)   
 \frac{\sqrt{Q^2(K^2+t)}}{8 \beta \sqrt{\pi} M 
\sqrt{x_{bj}^{in}x_{bj}^{fin}}E_0} \nonumber \\
&&\times\left[e^{-\frac{M^2}{2\beta^2}(\frac{E_0}{M} - x^{in}_{bj})^2}
e^{\frac{M^2(x_{bj}^{in} - x_{bj}^{fin})^2}{4 \beta^2}}
e^{-\frac{M^2}{2\beta^2}(\frac{E_0}{M} - x^{fin}_{bj})^2}\right]
e^{-(\frac{\vec{\Delta}^2}{4 \beta^2})} \ ,
\label{factored}
\eea
and by steps analogous to section III, the structure function
can be expressed as

\bea
F_2(x_{bj}^{in},x_{bj}^{fin},t) & =&   (e_1^2 + e_2^2)
x_{bj}^{in}x_{bj}^{fin} \frac{M^2}{\beta \sqrt{\pi}E_0} \nonumber \\
&&\times\left[e^{-\frac{M^2}{2\beta^2}(\frac{E_0}{M} - x^{in}_{bj})^2}
e^{\frac{M^2(x_{bj}^{in} - x_{bj}^{fin})^2}{4 \beta^2}}
e^{-\frac{M^2}{2\beta^2}(\frac{E_0}{M} - x^{fin}_{bj})^2}\right]
e^{-(\frac{\vec{\Delta}^2}{4 \beta^2})} \ ,
\label{factored2}
\eea
which is to be compared with Eq.~(\ref{eq:17}) for the forward case.

Thus Eq.~(\ref{factored2}) can be written as
\be
F_2(x_{bj}^{in},x_{bj}^{fin},t) 
\sim  \sqrt{ F_2(x_{bj}^{in}) \times  F_2(x_{bj}^{fin})}  
\times F_{00}(\vec{\Delta}_T) \ .
\ee
The factor 
$F_{00}(\vec{\Delta}_T)$ is
dependent
on $x_{bj}^{in}$, $x_{bj}^{fin}$ and $t$. 
When $x_{bj}^{in} - x_{bj}^{fin} (\equiv \xi)  \to 0$
 this form reduces to the parton density distribution modulated by the
{\it transverse} momentum transfer (compare Ref.~\cite{burk}).
However, we can restore the $F_{el}(t)$ by recognising the full import of
Eqs.~(\ref{factored}) and (\ref{factored2}).

Comparison with the forward scattering Eq.~(\ref{eq:17}) 
shows that the first and third exponentials in 
the square bracket of Eq.~(\ref{factored2}) are effectively the
{\it amplitudes} for finding a parton at $x_{bj}^{in}$ 
in the initial hadron and at $x_{bj}^{fin}$ in the final hadron. 
From Eq.~(\ref{factored2}) the term outside the square bracket, 
$e^{-(\vec{\Delta}^2/4 \beta^2)}$,
is seen to be the elastic form factor for the composite state $A$, 
$F_{el}(t) \equiv F_{00}(t)$, 
which arises because the
initial and final states are effectively scattered elastically 
with invariant four momentum transfer squared 
$t \equiv -\vec{\Delta}^2$. The central term in the square bracket,
$e^{M^2(x_{bj}^{in} - x_{bj}^{fin})^2/4 \beta^2}$, appears to be
a specific property of the sum over intermediate states in the
dual model and merits some discussion.

As Fig.~\ref{fig:(4)} schematically illustrates, 
we explicitly included coherent 
intermediate ``resonant" states.
 Although we recover the leading twist
of the quasi-free parton model, nonetheless confinement is present 
and imposing itself throughout, in particular in the intermediate
hadron state for which a parton has entered with $x_{bj}^{in}$ 
and departed with $x_{bj}^{fin}$.
 If we considered the limit $\beta \to 0$, recovering the most naive form 
 of a quasi-free independent
parton model, with no memory of confinement and no non-trivial 
excitation spectrum in the intermediate
state, this term will vanish (unless $x_{bj}^{in} = x_{bj}^{fin}$), 
as will the entire
amplitude. As mentioned in section III, $\beta \neq 0$ leads to 
a physical excitation gap, 
which both smears the distributions
and leads to a finite overlap when $x_{bj}^{in} \neq x_{bj}^{fin}$.

It is $\beta \neq 0$ that enables duality via this intermediate state overlap. 
Specifically,
the basic photon-parton ($Q$) scattering 
$\gamma(q) Q \to \gamma(k) Q$ has a non-zero 
cross section, as does the hadronic
process $\gamma A \to A^* \to \gamma(k)A$ via a specific intermediate 
state $A^*$. However, the duality is
non-trivial. If confinement is ``hidden" such that $\beta \to 0$, 
then although   $\gamma(q) Q \to \gamma(k) Q$
exists, its embedding in the  initial and final composite states will 
vanish due to the 
misalignment of momenta of the struck parton relative to that 
of the spectator(s). In this case the duality
is still realised because all the $A^*$ states become degenerate 
and destructively interfere unless $x_{bj}^{in} = x_{bj}^{fin}$.

Thus, the structure in Eq.~(\ref{factored}) 
suggests that the non-forward structure function  
may generalize to a factorisation between:

(i) the hadron-parton distribution amplitudes; 

(ii) a ``quark-in, quark-out"
term associated with the intermediate coherent state;

(iii) the invariant momentum transfer $t$.

We can restore the appearance of $F_{00}(t)$ by recognising that
contribution (ii) describes the longitudinal part of the momentum 
transfer being ``shared" in a two-step process - 
excitation and decay of the intermediate coherent state. 
In effect it is an indication of so-called 
$\xi$-dependence discussed 
in the literature~\cite{Ji-Mel-Song,rad-prd-98,petrov,goeke},
arising from the ``memory" of coherent confinement 
in the (sum over) intermediate
states. 
\noindent Thus Eq.~(\ref{factored2}) may be generalised to,
\bea
F_2(x_{bj}^{in},x_{bj}^{fin},t) 
&\equiv & \sqrt{ F_2(x_{bj}^{in}) \times 
\frac{1}{F_{00}^2[M^2(\xi)^2]}\times F_2(x_{bj}^{fin})} 
\times F_{00}(t)  \nonumber \\ 
&\sim & \sqrt{ F_2(x_{bj}^{in}) \times  F_2(x_{bj}^{fin})}  
\times F_{00}(\vec{\Delta}_T) \ .
\label{factor}
\eea

Equation~(\ref{factor}) has the advantage of separating out
the form factors 
governed by the external composite system [$F_{00}(t)$] and 
internal ones. As discussed in the previous paragraph, 
such a form also explains the onset of the ``quark-hadron" duality 
in DVCS.

It is also interesting to note that the original non-forward
structure function automatically satisfies the positivity constraint
of Ref.~\cite{pire-99}.
In Eq.~(\ref{rldvcs1}), one can treat the non-forward structure 
function as a product of two forward ones with incoming photon momenta 
$q$ and $k$ associated with a rotation function.
Apart from the rotation function, the positivity bound for the 
non-forward structure function thus can be realized.
Equation~(\ref{factor}) possesses the same feature, in which the
positivity bound gives an inequality as follows:
\be
F_2(x_{bj}^{in},x_{bj}^{fin},t)\le 
\sqrt{F_2(x_{bj}^{in}) \times  F_2(x_{bj}^{fin})} \ .
\ee

An alternative way to write Eq.~(\ref{factored})
is
\bea
F_L(x_{bj}^{in},x_{bj}^{fin},t,Q^2) & =& (e_1^2 + e_2^2)\frac{\sqrt{Q^2(K^2+t)}}
{8 \beta \sqrt{\pi} M \sqrt{x_{bj}^{in}x_{bj}^{fin}}E_0} \nonumber\\
&&\times\left[
e^{-\frac{M^2}{\beta^2}
\left(\frac{E_0}{M} - \frac{x^{in}_{bj}+ x^{fin}_{bj} }{2}\right)^2} 
\right] 
e^{-(\frac{\vec{\Delta}^2}{4 \beta^2})} \ ,
\eea
which can be generalised to
\be
F_L(x_{bj}^{in},x_{bj}^{fin},t,Q^2) \equiv F_L(x,t,Q^2) 
\times F_{00}(t) \times 
\left[1 - \left(\frac{\xi}{x}\right)^2\right] \ ,
\label{eq:31}
\ee
where we have defined:
\be
x \equiv (x_{bj}^{in} + x_{bj}^{fin})/2 ; \ 
\xi \equiv (x_{bj}^{in} - x_{bj}^{fin})/2 \ .
\label{symx}
\ee
Accordingly, in this case the structure function can be written as
\be
F_2(x,\xi,t)=(e_1^2+e_2^2)
\frac{(x-\xi)(x+\xi)}{x^2}F_2(x)F_{el}(t) \ ,
\label{3rd-fact}
\ee
where $F_2(x)$ is the structure function for the forward
scattering. 

The reduction of Eq.~(\ref{3rd-fact}) to forward
Compton scattering is obvious. 
If there were no intermediate coherent state, 
we expect that the physics would be
symmetric in $x_{bj}^{in}$ and $x_{bj}^{fin}$,
through which the forward Compton scattering ($x_{bj}^{in}=x_{bj}^{fin}$)
corresponds to the kinematics, $x=x_{bj}^{in}=x_{bj}^{fin}$ and $\xi \to 0$.
We conjecture that if the intermediate coherent state were absent, 
as in the incoherent parton model description,
a plausible factorisation between the overall $F_{el}(t)$ and parton 
effective probability distribution, symmetric
in $x_{bj}^{in}$ and $x_{bj}^{fin}$ would be
\be
F_2(x_{bj}^{in}=x_{bj}^{fin},t) 
\equiv F_2(x,\xi=0, t) =  F_2(x) 
\times F_{el}(t) \ .
\label{averagefactor}
\ee
Note that $x \equiv (x_{bj}^{in} + x_{bj}^{fin})/2 $
and so this contains Eq.~(\ref{xover2}) 
as a special case in the approximations that were
used there, whereby $x_{bj}^{fin} \to 0$ and, as remarked earlier, differing
from a commonly used {\it ansatz}. 

In the Gaussian wavefunction model, 
the equivalence between Eq.~(\ref{factor}) and (\ref{3rd-fact})
 is exact.
In Eq.~(\ref{averagefactor}) the non-forward
amplitude effectively becomes an ``average" of the forward distributions,
modulated by the kinematic factor 
$[1 - (\xi/x)^2]$ where $\xi\to 0$ in the forward case. 
It remains to be investigated
whether this analytic equivalence between Eqs.~(\ref{factor}) 
and (\ref{3rd-fact})
is an artifact of the Gaussian distributions and
the restriction to $F_L$ with spinless constituents. 
The latter in particular needs study as 
the imaginary part of DVCS is measurable 
in electron scattering from a polarised target, 
and with spinless constituents this is manifestly
a non-leading effect.

\section{Phenomenology}

In this paper we have constructed a model for
composite systems consisting of two spinless particles.
The functional relations between the forward and non-forward
distributions, and the elastic form factor of the composite system,
are summarised in  Eqs.~(\ref{factor}) and (\ref{3rd-fact}).
For phenomenology we want to investigate the implications
of these factorisations for the non-diffractive 
structure functions of the nucleon.
We adopt a dipole elastic form factor $F_{el}(t) =1/ (1-t/\lambda)^2$,
where $\lambda=0.7$ (GeV/c)$^2$ is the empirical energy scale, 
and choose $|t| = 1$ (GeV/c)$^2$, whereby
the factorisation
of Eq.~(\ref{factor}) implies that the
non-forward parton distribution will have the form:
\be
F_2(x_i,x_f,t) =\sqrt{F_2(x_i) F_2(x_f)} 
\frac{F_{el}(t)}{1+(x_i-x_f)^2 M^2/\lambda}  \ .
\label{phenom-1}
\ee
Moreover, for more realistic predictions relevant to a three-body
constituent system, we adopt the commonly used parametrisation
$F_2(x)=x(1-x)^3$. The results are displayed in Fig.~\ref{fig:(5)}.
As $x_i (x_f) < 1$, one has $(x_i - x_f)^2 <<1$ and the $F_2(x_i,x_f,t)$ 
is almost symmetric in
$x_i$ and $x_f$. Hence for fixed $x_f$ the figure shows a rise and fall as 
a function of $x_i$
similar to that observed for the ``usual" forward structure function,
and to a bag model study by Ji {\it et al}~\cite{Ji-Mel-Song}. 
However, there are significant dynamical differences
with that model: we find the distribution exhibits a sensitivity to $x_f$, 
which is not apparent in~\cite{Ji-Mel-Song}.
In contrast to Ref.~\cite{Ji-Mel-Song}, 
in which the origins of the $\xi$ insensitivity are obscure~\cite{jijphys},
 the systematics of the distribution in the present model,
 especially its $x_f$ dependence, 
are more transparent.

For comparison, we also investigate the factorisation of Eq.~(\ref{3rd-fact})
using the same quark counting rule parametrization as input.
The original form only applies in the limit $x_{bj}^{in}>x_{bj}^{fin}$
and so it is not clear whether its generalisation should be taken seriously
out of that region. 
However, if we take the generalized form 
to apply throughout the entire physical range where the variable $x$ and $\xi$
each have $-1\le x\le 1$ and $\xi \ge 0$,
we obtain some interesting features.
As shown in Fig.~\ref{fig:(6)}, with $|t|=1$ (GeV/c)$^2$,
the distribution function 
is positive at $x>\xi$, which is the region dominated by the constituent (quark) distributions.
The distribution
function exhibits significant 
sensitivity to $\xi$. 
The region $x>>\xi$, which corresponds to $Q^2\sim |t|$,
is the place where resonance effects could play a role if $|t|$ is small.
As is to be expected with our quark counting rule parameterization,
the largest probabilities of the quark distribution
occur at $\sim 1/3$. 
A sign change occurs at $x=\xi$, where $Q^2>> |t|$. 
In the region
$x<\xi$ where $K^2+t > Q^2$, the partons have
negative momentum fraction $x_{bj}^{fin}=x-\xi<0$; this may be interpreted as
due to antiquark $\bar{q}$ which can play a role in this kinematic region.
Interestingly, the interpretation of this generalisation
is consistent with that in Ji's frame (see e.g. discussions 
in Ref.~\cite{rad-99}). 
One can see that with $\xi\to 0$, which corresponds to 
larger $Q^2$ with a fixed $t$, 
this factorisation succeeds in reproducing the forward Compton scattering.
A scaling behavior is also observed.
However, notice that 
a fast crossover occurs at $x=0$ if $x<\xi$, where the usual quark
density becomes infinite. 
This feature could imply that the physical region of this generalisation
form only make sense at $x> \xi$~\cite{jijphys,rad-99}.

\section{Conclusion and discussion}

We have an explicit model, in which both forward 
and non-forward Compton scattering have been investigated.
The model also exhibits scaling. Although this model is far from reality,
we may still draw intuition from it as to how the physical DVCS and 
related processes may behave.

For forward scattering we have extended the work of Ref.~\cite{IJMV}
to constituents with arbitrary charges and given an explicit dynamical 
model realising the general picture outlined by Close and Isgur~\cite{ClIs01}.
For the first time we have also determined the implications
of such models for the non-forward Comption scattering.
We stress that this is at best only a pedagogic picture
due to its emphasis on spinless constituents.

The model showed scaling behavior explicitly for the forward scattering
and also satisfied well known sum rules. This also has been extended to 
the non-forward case. We find here too that scaling is predicted and 
explicit $\xi$ dependence is also seen.

The behaviour of the Compton amplitude at $90^\circ$ showed 
how the effective simple
$s^{-n}$
dependence, previously derived from counting rules~\cite{brodsky}, arises.
For us it is directly driven by the dependence of $F_{el}(t)$ once our
Gaussian forms are generalised to phenomenological form factors.
We found that the degeneracy among states with a common $N (=L +2k)$
but different $L$ causes a destructive interference among all but the
elastic Born term. Thus the $s^{-n}$ behaviour, in this interpretation,
is effectively dominated by elastic scattering in the direct channel.
Away from $90^\circ$ in the model this interference is no longer exact. Indeed,
in the real world where this $N$-$L$ degeneracy is broken (e.g. by spin
dependent effects arising from one-gluon exchange), oscillations around
the overall average smooth $s^{-n}$ are predicted rather naturally. 
It is therefore interesting that such effects are qualitatively
evident in the data~\cite{oscil}. A quantitative investigation of this will
be reported elsewhere.

Our work is restricted to the imaginary part of DVCS. 
This in turn is measurable in electron scattering with polarized target.
This exposes the limit of our spinless model.
It is the longitudinal response functions that are leading in such a model
and exhibit interesting factorisations
(e.g. Eq.~(\ref{factor}) and ~(\ref{3rd-fact})). We expect that such 
relations will survive for investigations with more realistic models.
However, it is perhaps interesting to note that the factorizations exhibited 
in the present model are novel and different from some popular 
{\it ansatz} used in the literature~\cite{vand-98}.

\acknowledgements

This work is supported, in part, by
the European Community Eurodafne,
contract CT98-0169, and the U.K. Engineering and Physical 
Sciences Research Council (Grant No. GR/M82141).
We are grateful to W. Melnitchouk, W. Van Orden, S. Jeschonnek, M. Guidal,
and N. Isgur for
discussions about their work and to J. Dudek for comments on the manuscript.

\section*{Appendix 1}

For a given excited level $N$, there are degenerate states 
with $L=N,N-2,\cdots 0(1)$ for $N=even(odd)$.
Whereas in the forward scattering we dealt immediately 
with $\sum_N |F_{0N}(\vec{q})|^2$, here we must first
sum over the various $L$ states within a given $N$, thus

\[
\sum_N\left( \sum_{L=0(1)}^N {\cal F}_{0N}^{(L)}(\vec{k})  
{\cal F}_{N0}^{(L)}(\vec{q}) d_{00}^L(\theta)  \right) \ ,
\]
where $\theta$ is the relative angle between the
incoming and outgoing photon momentum, 
i.e. $\vec{k}\cdot\vec{q}=kq\cos(\theta)$ (in what follows
$q$ and $k$ denote the magnitudes of the three-vector momenta
$\vec{q}$ and $\vec{k}$, respectively).
In the potential model of Ref.~\cite{IJMV}, after sum over $L$
the above expression reduces to
\be
\label{sum-over-n}
\sum_{N} F_{0N}(\vec{q})F_{N0}(\vec{k}) = \sum_{N} 
\frac{1}{N!}\left(\frac{\vec{q} \cdot \vec{k}}{2\beta^2}\right)^N \exp
(-\frac{\vec{q}^2 + \vec{k}^2}{4\beta^2}) \ .
\ee

As illustration we show the first non-trivial case for $N=2$. 
Here, for $L=2$
\[
{\cal F}_{02}^{(2)}(\vec{k})  {\cal F}_{20}^{(2)}(\vec{q}) 
d_{00}^2(\theta)=\frac{1}{3} 
\left(\frac{3\cos^2(\theta) -1}
{2}\right)\left(\frac{kq}{2\beta^2}\right)^2  
e^{-(\vec{q}^2 + \vec{k}^2)/4\beta^2} \ ,
\]
and for $L=0$ 
\[
{\cal F}_{02}^{(0)}(\vec{k})  {\cal F}_{20}^{(0)}(\vec{q}) 
d_{00}^0(\theta)=\frac{1}{6}  \left(\frac{kq}{2\beta^2}\right)^2 
e^{-(\vec{q}^2 + \vec{k}^2)/4\beta^2} \ .
\]
Thus
\[
F_{02}(\vec{q})F_{20}(\vec{k}) = 
{\cal F}_{02}^{(2)}(\vec{k})  {\cal F}_{20}^{(2)}(\vec{q}) d_{00}^2(\theta)+ 
{\cal F}_{02}^{(0)}(\vec{k})  {\cal F}_{20}^{(0)}(\vec{q}) d_{00}^0(\theta) \ ,
\]
which is the $N=2$ component in Eq.~(\ref{sum-over-n}).

That this result generalises is readily seen in the Cartesian basis where
 the sum over resonances for non-forward Compton scattering
in this relativistic model.
We start with the general form
\bea
\label{non-forward-trans}
M&=& \delta (\vec{q}+\vec{P}_i-\vec{k}-\vec{P}_f)
\sum_N\langle\Psi_0(\vec{r})|[e_1e^{-i\vec{k}\cdot\vec{r}/2}
+e_2e^{i\vec{k}\cdot\vec{r}/2}]|\Psi_N\rangle\nonumber\\
&&\times\langle \Psi_N|
[e_1e^{i\vec{q}\cdot\vec{r}/2}
+e_2e^{-i\vec{q}\cdot\vec{r}/2}]|\Psi_0(\vec{r})\rangle \ ,
\eea
where $\Psi_N=\psi_{n_x}\psi_{n_y}\psi_{n_z}$ 
is the harmonic oscillator wave function, and $N=n_x+n_y+n_z$
is the main quantum number.
The one dimension harmonic oscillator, e.g. the $z$ component 
has the expression
\be
|\psi_{n_z}(z)\rangle=\left[\frac{\beta}{2\sqrt{\pi}2^{n_z} n_z!}\right]^{1/2}
e^{-\frac 18 \beta^2 z^2} H_{n_z}(\beta z/2) \ ,
\ee
where $H_{n_z}(\beta z/2)$ is the Hermite polynomial and
the orthogonal relation is,
\be
\int_{-\infty}^{+\infty}
H_{n}(x)H_{m}(x)e^{-x^2} dx=\sqrt{\pi} 2^n n!\delta_{nm} \ .
\ee
For the excitation process we fix the incoming photon momentum 
to be colinear with
that of the target, and refer to it here as the $z$ direction. 
Only the $z$ compoment $\psi_{n_z}$ will be excited. 
We have
\be
\langle \psi_{n_z}(z)|e^{i qz/2}|\psi_0(z)\rangle 
=\frac{1}{\sqrt {n_z !}}
\left(\frac{iq}{\sqrt{2}\beta}\right)^{n_z} e^{-q^2/4\beta^2} \ .
\ee

The excited intermediate state will emit a photon with momentum $\vec{k}$
and fall to the ground state.
The transition amplitude thus can be expressed as
\be
\langle \psi_0(x)\psi_0(y)\psi_0(z) 
|e^{-i(k_x x+k_y y+ k_z z)/2}
|\psi_0(x)\psi_0(y)\psi_{n_z}(z)\rangle \ .
\ee
Following the same strategy, we can explicitly derive,
\be
\langle \psi_0(z)|e^{-ik_z z/2} |\psi_{n_z}(z)\rangle 
=\frac{1}{\sqrt{n_z !}}
\left(-\frac{ik_z}{\sqrt{2}\beta}\right)^{n_z} e^{-k_z^2/4\beta^2} \ ,
\ee
for the $z$ component. The $x$ and $y$ components are 
\bea
\langle \psi_0(x)|e^{-ik_x x/2} |\psi_0(x)\rangle 
& =& e^{-k_x^2/4\beta^2} \ , \nonumber\\
\langle \psi_0(y)|e^{-ik_y y/2} |\psi_0(y)\rangle 
& = & e^{-k_y^2/4\beta^2} \ .
\eea

Thus, we obtain
\be
\langle \psi_0(x)\psi_0(y)\psi_0(z) 
|e^{-i\vec{k}\cdot\vec{r}/2}
|\psi_0(x)\psi_0(y)\psi_{n_z}(z)\rangle 
=\frac{1}{\sqrt{n_z !}}
\left(-\frac{ik\cos\theta}{\sqrt{2}\beta}\right)^{n_z} 
e^{-\vec{k}^2/4\beta^2} \ ,
\ee
where $k_z=k\cos\theta$, 
and $\theta$ is the angle between the $\vec{k}$ and $z$ direction 
(the direction of $\vec{q}$).

The total transition of Eq.~(\ref{non-forward-trans})
can be then expressed as
\bea
M&=&
\delta (\vec{q}+\vec{P}_i-\vec{k}-\vec{P}_f)
\sum_{N=0}^{\infty} e^{-(\vec{k}^2+\vec{q}^2)/4\beta^2}\nonumber\\
&&\times\left[(e_1^2+e_2^2)\frac{1}{N !}
\left(\frac{ikq\cos\theta}{2\beta^2}\right)^N
+2e_1 e_2\frac{1}{N !}\left(-\frac{ikq\cos\theta}{2\beta^2}\right)^N\right] \ ,
\eea
where the main quantum number $N=n_z$ 
in this transition (which in general will include
degenerate states with $L=N, N-2 \cdots$). 
The above deduction gives the origin of Eqs.~(\ref{sum-over-n})
and ~(\ref{rldvcs1}).

\section*{Appendix 2}

In the previous Sections,
the term proportional to $e_1e_2$ has not been discussed in detail.
Here, we shall show that this term would vanish 
in the limit of $N\to\infty$ for the non-forward Compton scattering process.

The positive energy term proportional to $e_1e_2$ in the structure function 
(Eq.~(\ref{eq:r2a})) is,
\be
2e_1e_2\sum_{N=0}^{\infty}
\frac{\sqrt{\nu^2+Q^2}}
{4E_0\beta^2}\frac{1}{ N!}\left(-\frac{\vec{q}\cdot\vec{k}}{2\beta^2}\right)^N
e^{-\frac{\vec{q}^2+\vec{k}^2}{4\beta^2}} \delta(E_N-E_0-\nu) \ .
\ee
Correspondingly, the term proportional to $e_1e_2$ in 
the scaling function is,
\be
R_L(\nu,\vec{q},\vec{k},t)=
2e_1e_2\sum_{N=0}^{\infty}
\frac{|\vec{q}|\sqrt{\nu^2+Q^2}}
{4E_0\beta^2}\frac{1}{ N!}\left(-\frac{\vec{q}\cdot\vec{k}}{2\beta^2}\right)^N
e^{-\frac{\vec{q}^2+\vec{k}^2}{4\beta^2}} \delta(E_N-E_0-\nu) \ ,
\ee
where $|\vec{q}|^2=2\beta^2 N$. 
Compared to the term proportional to $(e_1^2+e_2^2)$,
a factor $(-1)^N$ arises from the sum over the resonances.
We therefore separately consider the $N=even$ and $odd$ terms
which gives,
\bea
R_L(\nu,\vec{q},\vec{k},t)&\equiv& R_L^{even}+R_L^{odd}\nonumber\\
&=&2e_1e_2 e^{-\frac{\vec{q}^2+\vec{k}^2}{4\beta^2}}
\left[\sum_{n=0}^{\infty}
\frac{|\vec{q}|\sqrt{\nu^2+Q^2}}
{4E_0\beta^2}\frac{1}{ (2n)!}
\left(\frac{\vec{q}\cdot\vec{k}}{2\beta^2}\right)^{2n} 
\delta(E_{2n}-E_0-\nu) \right. \nonumber\\
&&\left. -\sum_{n=0}^{\infty}
\frac{|\vec{q}|\sqrt{\nu^2+Q^2}}
{4E_0\beta^2}\frac{1}{ (2n+1)!}
\left(\frac{\vec{q}\cdot\vec{k}}{2\beta^2}\right)^{2n+1} 
\delta(E_{2n+1}-E_0-\nu) \right] \ .
\eea

Following the analogous steps as outlined
for the term of $(e_1^2+e_2^2)$ 
in Sec. III, the factorial 
part can be expanded the same way by using the Stirling's formula.
After some tedious, but essentially the same algebra, 
we obtain,
\bea
& & R_L^{even}+R_L^{odd}\nonumber\\
&=&\frac{2e_1e_2}{4\beta\sqrt{\pi} E_0}e^{-\frac{M^2}{\beta^2}
(\frac{E_0}{M} - x_{bj}+\frac{\vec{q} \cdot \vec{\Delta}}{2 M \nu})^2}
e^{-(\frac{\vec{\Delta}^2}{4 \beta^2})} \nonumber\\
&&\times [\delta(E_{2n}-E_0-\nu) -  \delta(E_{2n+1}-E_0-\nu) ] \ ,
\eea
where the $\delta$ functions are 
shown explicitly
because we assume that
the photon energy $\nu$ always satisfies the condition that
the excited states are $2n$ for $R_L^{even}$ and $2n+1$ for $R_L^{odd}$. 
Consequently, taking the limit $n\to\infty$, 
$E_{2n}=E_{2n+1}$ can be satisfied, which thus leads 
to $R_L^{even}+R_L^{odd}=0$. Namely, the twist term vanishes.
Similar investigation for the energy negative solution
results in the same conclusion.


\begin{figure}
\begin{center}
\epsfig{file=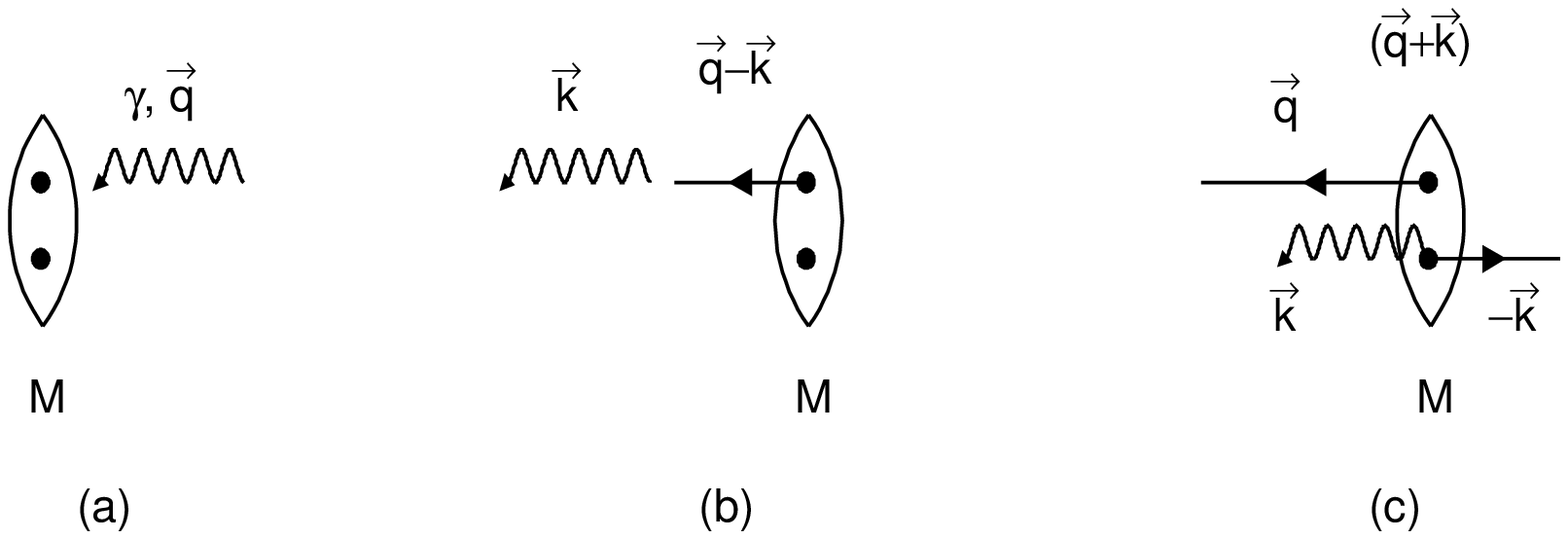,height=6.0cm,width=10.0cm}
\end{center}
\caption{ 
Schematic diagrams of DVCS for a two-scalar-constituent system.  }
\protect\label{fig:(1)}
\end{figure}
\begin{figure}
\begin{center}
\epsfig{file=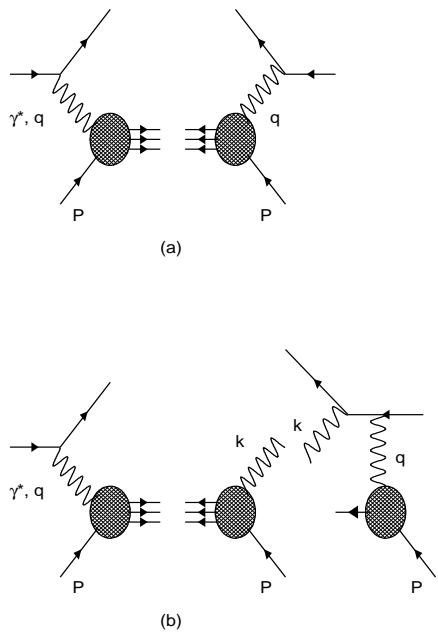,height=12.0cm,width=8.0cm}
\end{center}
\caption{ 
Schematic diagrams of forward and non-forward DVCS. }
\protect\label{fig:(2)}
\end{figure}
\begin{figure}
\begin{center}
\epsfig{file=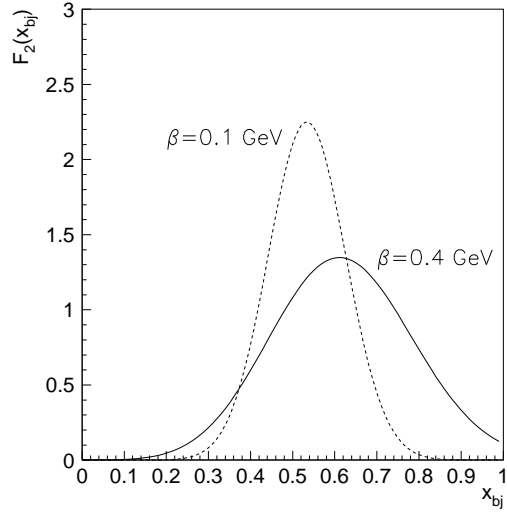,height=8.0cm,width=8.0cm}
\end{center}
\caption{ 
Structure function $F_2(x_{bj})$ without
the charge factor. $\beta$ denotes the linear potential strength. }
\protect\label{fig:(3)}
\end{figure}
\begin{figure}
\begin{center}
\epsfig{file=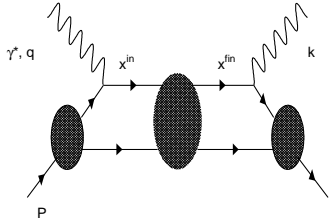,height=8.0cm,width=8.0cm}
\end{center}
\caption{ 
Schematic diagram for the quark-hadron duality 
in the non-forward Compton scattering. }
\protect\label{fig:(4)}
\end{figure}
\begin{figure}
\begin{center}
\epsfig{file=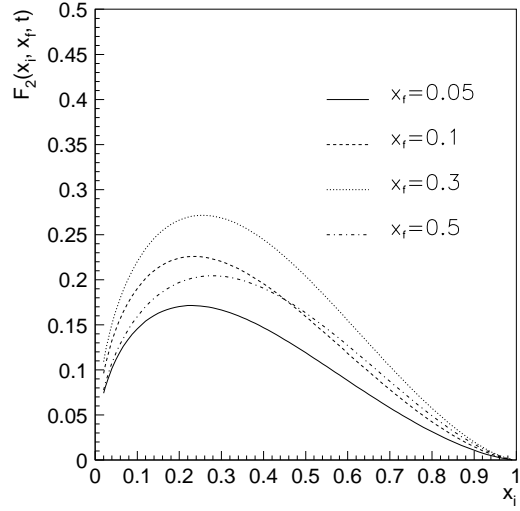,height=8.0cm,width=8.0cm}
\end{center}
\caption{ 
Non-forward structure function following the phenomenology
of Eq.~(\ref{phenom-1}) with $t=-1$ (GeV/c)$^2$ fixed.
As $x_f/x_i \sim -t/Q^2$, $Q^2$ increases with increasing $x_i$. }
\protect\label{fig:(5)}
\end{figure}
\begin{figure}
\begin{center}
\epsfig{file=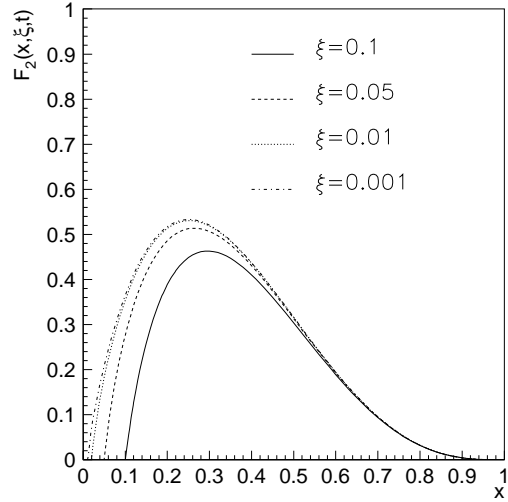,height=8.0cm,width=8.0cm}
\end{center}
\caption{
Non-forward structure function following factorisation of Eq.~(\ref{3rd-fact})
with $t=-1$ (GeV/c)$^2$ fixed. Note that $t \to 0$ as $x \to \xi$. }
\protect\label{fig:(6)}
\end{figure}

\end{document}